\documentclass{optica-article}
\DeclareUnicodeCharacter{2212}{-}
\DeclareUnicodeCharacter{03BC}{$\mu$}
\journal{opticajournal} 

\articletype{Research Article}

\usepackage{braket}
\usepackage{lineno}

\begin{document}

\title{Extraction of coherence times of biexciton and exciton photons emitted by a single resonantly excited quantum dot under controlled dephasing}

\author{Jaewon Lee,\authormark{1} Charlie Stalker,\authormark{1}, Loris Colicchio\authormark{1}, Fernando Redivo Cardoso\authormark{2,1}, Jan Seelbinder\authormark{1}, Sven H\"{o}fling\authormark{3}, Christian Schneider\authormark{4}, Celso J. Villas-Boas\authormark{2}, Ana Predojevi\'{c}\authormark{1}\authormark{*}}

\address{\authormark{1}Department of Physics, Stockholm University, 10691 Stockholm, Sweden\\

\authormark{2}Departamento de F\'{i}sica, Universidade Federal de S\~{a}o Carlos, 13565-905 S\~{a}o Carlos, S\~{a}o Paulo, Brazil\\

\authormark{3}Technische Physik, Physikalisches Institut and W\"urzburg-Dresden Cluster of Excellence ct.qmat, Universit\"at W\"urzburg, Am Hubland, D-97074 W\"urzburg, Germany\\

\authormark{4}Institut of Physics, University of Oldenburg, D-26129 Oldenburg, Germany}

\email{\authormark{*}ana.predojevic@fysik.su.se} 

\begin{abstract*} 
The visibility of two-photon interference is limited by the indistinguishability of the photons. In the cascaded emission of a three-level system, such as a single quantum dot, the indistinguishability of each photon in the pair is primarily affected by two main factors: the temporal correlation between paired photons and dephasing. Investigating the individual effects of these factors on photon indistinguishability is challenging, as both factors affect it simultaneously. In this study, we investigate the temperature-dependent two-photon interference visibility of the biexciton and exciton photons emitted from a single quantum dot under two-photon resonant excitation, while keeping temporal correlation between the paired photons intact. Finally, we simultaneously extract the coherence times of the biexciton and exciton photons as a function of temperature.

\end{abstract*}

\section{Introduction}

Photons are versatile carriers of information due to their low interaction with the surrounding environment, making them well-suited for quantum communication protocols such as quantum key distribution \cite{bennett_quantum_2014,ekert_quantum_1991}. However, many quantum information protocols, such as quantum teleportation \cite{bennett_teleporting_1993} and entanglement swapping \cite{yurke_einstein-podolsky-rosen_1992}, require an effective photon-photon interaction, which is fundamentally challenging because of the same low interaction property of photons. Two-photon interference on a beamsplitter, also known as Hong-Ou-Mandel (HOM) interference \cite{hong_measurement_1987}, provides a means of inducing photon-photon interactions. Therefore, achieving high visibility in HOM interference is a critical step in advancing photon-based quantum technologies. 

Quantum dots serve as a versatile platform for the on-demand generation of photon pairs through coherent control \cite{jayakumar_deterministic_2013, huber_semiconductor_2018}. This on-demand generation of photon pairs makes quantum dots attractive for quantum photonics applications. However, despite this advantage, quantum dots exhibit low HOM interference visibility in photon pair generation relying on a cascaded emission, whereas they exhibit very high HOM interference visibility when used as single-photon sources \cite{ding_-demand_2016}. This limitation hinders the implementation of quantum technologies that rely on high HOM visibility of single photon pairs. One of the main factors reducing HOM interference visibility is the temporal correlation between paired photons due to the cascaded emission \cite{franson_bell_1989,huang_correlations_1993, simon_creating_2005}. In a single quantum dot, photon pairs are emitted via cascaded emission in a three-level system, where the biexciton photon is always emitted before the exciton photon. This time ordering is reflected in the wavefunction of the biexciton and exciton photons as follows \cite{huang_correlations_1993, simon_creating_2005}:

\begin{equation}\label{single cascade wavefunction}
    \psi(t_b,t_x) = 2 \sqrt{\Gamma_b \Gamma_x} e^{-\Gamma_b t_b} \Theta(t_b) e^{-\Gamma_x (t_x - t_b)} \Theta(t_x - t_b),
\end{equation}
where  $\Gamma_{i}$ and  $t_{i}$ represent the decay rate and emission time, respectively, and the subscripts $b$ and $x$ correspond to the biexciton and exciton, respectively. In particular, the Heaviside function $\Theta(t_{x} - t_{b})$ imposes the time-ordered emission, creating a temporal correlation between the biexciton and exciton photon pair. This temporal correlation reduces the purity of photonic states, with the extent of reduction depending on the decay rates of the biexciton and exciton. The purity of exciton photon is expressed as follows \cite{huang_correlations_1993, simon_creating_2005}:

\begin{equation}\label{trace reduced rho}
    Tr(\rho_x^2) = \frac{\Gamma_b}{\Gamma_b + \Gamma_x},
\end{equation}
which leads to a reduction in the visibility of HOM interference.

The other key factor that affects HOM visibility is the coherence time ($T_2$) of the emitted photons. However, assessing the coherence times of photons from cascaded emission is not straightforward due to the interplay between coherence time and temporal correlation. For instance, in a two-level system, where the exciton is resonantly excited and emits a single photon, the coherence time can be estimated from HOM interference measurements \cite{unsleber_two-photon_2015,thoma_exploring_2016,varoutsis_restoration_2005}. In contrast, in a three-level system with cascaded emission, the HOM visibilities of the biexciton and exciton photons depend
not only on their individual coherence times, but also on their temporal correlation. 

Due to this complexity, the effect of phonon-induced dephasing on the visibility of HOM interference and the coherence time has been reported only for the exciton photon \cite{thoma_exploring_2016,varoutsis_restoration_2005}, while the effect of phonon-induced dephasing on the coherence times of both the biexciton and exciton photons under resonant excitation of the biexciton has never been studied simultaneously. This motivates the exploration of HOM interference visibility and coherence times for the biexciton and exciton photons under conditions of constant temporal correlation and the two-photon resonant excitation scheme, while systematically modifying phonon-induced dephasing, which can be controlled by changing the temperature of the sample.

In this article, we show the correlation functions of HOM as the phonon-induced dephasing in a single quantum dot is varied, while keeping the temporal correlation effects constant. We utilized superconducting nanowire single-photon detectors (SNSPDs) to achieve high temporal resolution in HOM interference measurements. Using the sensor method \cite{cardoso_impact_2025, del_valle_theory_2012}, we extract coherence times of the biexciton and exciton photons and relate them to the HOM interference visibility. This allows us to isolate the impact of phonon-induced dephasing on HOM interference visibility and to observe how coherence times of the biexciton and exciton photons alone affect the correlation functions of HOM interference. 
\begin{figure}[t]
\centering
\includegraphics[width=0.9\linewidth]{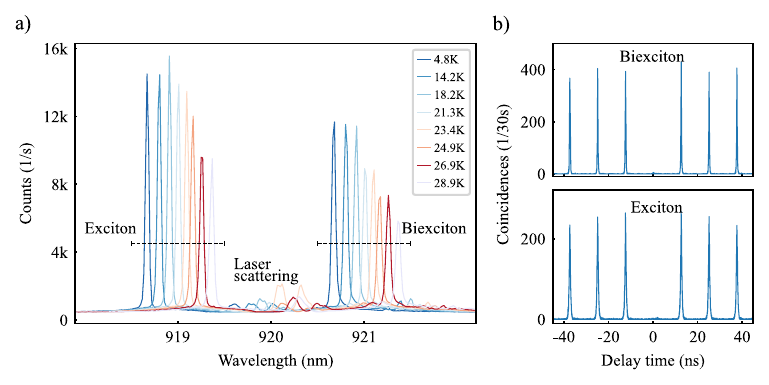} 
\caption{a) Emission spectra over different temperatures are shown, with the biexciton and exciton emission lines highlighted by dashed bands on the right and left, respectively. A redshift in the emission lines is observed as the temperature increases. Residual scattering from the excitation laser is observed between the emission lines under two-photon resonant excitation. b) Autocorrelation measurements for the biexciton (top) and exciton (bottom) photons at 4.8 K. The results yielded values of $g^{(2)}_{b}(0) = 0.0076(5)$ for the biexciton and $g^{(2)}_{x}(0) = 0.01543(12)$ for the exciton.}
\label{fig:spectra}
\end{figure}


\section{Methods}
\subsection{Experimental setup}

We used a single In(Ga)As quantum dot embedded in a self-aligned planar cavity with a bandwith of $\sim$5 nm, as described in \cite{maier_bright_2014, gines_time-bin_2021}. The sample was kept in a closed-cycle cryostat for the experiment. A Ti:Sapphire laser was used to excite the quantum dot, producing pulses with a repetition rate of 80 MHz. We tailored the excitation pulses using a pulse stretcher. The quantum dot was driven by the two-photon resonant excitation \cite{stufler_two-photon_2006,jayakumar_deterministic_2013}. The emitted photons were collected by an aspherical lens with a numerical aperture of 0.71. The excitation laser scattering was suppressed using a polarizer, a half-wave plate, and two notch filters. We spatially separated the biexciton and excition photons by directing them onto an optical grating and coupled them into single-mode fibers connected to SNSPDs. The experimental conditions, excluding the temperature, were kept constant during the measurements. Throughout the measurements, the temperature was set within a range of 4.8 K to 26.9 K. This change in temperature induced a red-shift in the emission lines, averaging 0.097 nm for the biexciton and 0.100 nm for the exciton between each set temperature, as shown in Fig. \ref{fig:spectra} a). Photon antibunching of the emission is confirmed by measuring the auto-correlation function, yielding values of $g^{(2)}_{b}(0) = 0.0076(5)$ for the biexciton and $g^{(2)}_{x}(0) = 0.01543(12)$ for the exciton at 4.8 K, as shown in Fig. \ref{fig:spectra} b).

\begin{figure}[t]
\centering
\includegraphics[width=0.8\linewidth]{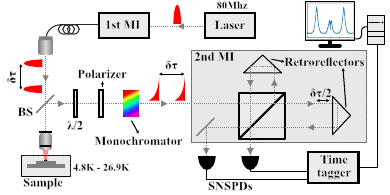} 
\caption{Schematic of the setup. The pulsed laser with 80 MHz repetition rate is directed into the first Michelson interferometer (1st MI), which generates two consecutive excitation laser pulses with approximately 3 ns delay time. These pulses sequentially excite a single quantum dot. To eliminate the polarization degree of freedom of single photons and thus remove photon distinguishability due to polarization, a half-wave plate and a polarizer are used. The emitted single photons impinge on a beamsplitter in the second Michelson interferometer (2nd MI), which is matched to the same delay time as the 1st MI. Coincidences between the two distinct output ports of the beamsplitter are recorded using SNSPDs and a time tagger.}
\label{fig:exp setup}
\end{figure}

For the measurement of the correlation function of HOM interference, we utilized a setup consisting of two Michelson interferometers \cite{santori_indistinguishable_2002}, shown in Fig. \ref{fig:exp setup}, both set to the same delay time between the short and long arms. The first interferometer was used to generate two consecutive excitation laser pulses with a delay of approximately 3 ns, which were used to excite the quantum dot to consecutively emit photon pairs. These emitted photons were then directed into the second interferometer, which served as the analyzer for two-photon interference. The second interferometer was set to the same delay time as the first, allowing two consecutively emitted photons to reach a beam splitter simultaneously, enabling HOM interference. We measured the correlation function by recording coincidences between two SNSPDs connected to the outputs of the analyzer. A detailed description of the experimental setup used for the measurements is given in \cite{huber_optimal_2015}.

\subsection{Theoretical model}
The correlation function of two single photons emerging from two output ports of the beamsplitter can be expressed in terms of the two different input modes of the beamsplitter, labeled 1 and 2, as follows \cite{legero_time-resolved_2003,kiraz_quantum-dot_2004}:

\begin{equation}\label{correlation}
    G^{(2)}_{HOM}(t,\tau)=\frac{1}{2}(n_{1}(t)n_{2}(t+\tau)-G^{(1)}_{1}(t,\tau)(G^{(1)}_{2}(t,\tau))^{*}),
\end{equation}
where $n_{i}$ refers to the photon number and $G^{(1)}_{i}(t,\tau)$ stands for the first-order correlation function on mode $i$. To calculate $G_{i}^{(1)}$ and the photon number $n_{i}$, we implement the sensor method \cite{cardoso_impact_2025, del_valle_theory_2012}. 
In this method, we introduce sensors consisting of two two-level systems, each weakly coupled to a three-level quantum dot system. These sensors, each corresponding to an emission frequency, are described by a Hamiltonian that includes their resonance frequencies and coupling to a three-level system of the quantum dot. A detailed theoretical explanation of the sensor method is provided in \cite{cardoso_impact_2025}.

Using this approach, we calculated the intensity of emission proportional to the photon number and $G^{(1)}$, then obtained $G^{(2)}_{HOM}(\tau)$ by integrating $G^{(2)}_{HOM}(t,\tau)$ over $t$. We simulated $G^{(2)}_{HOM}(\tau)$, using the experimentally determined lifetimes and coherence times of the biexciton and exciton which are in a range of selected values. By comparing the measured correlation functions of HOM interference with the theoretical predictions of $G^{(2)}_{HOM}(\tau)$, convolved with the system response, we extracted the coherence times of the biexciton and exciton \cite{del_valle_theory_2012,cardoso_impact_2025}.

\begin{figure}[h!t]
    \centering 
    \includegraphics[width=0.7\linewidth]{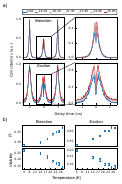}
    \caption{a) Experimentally measured correlation functions of HOM interference at different temperatures are shown at the top and bottom (left column) for the biexciton and exciton photons, respectively. The right column displays zoomed-in views of the central peaks, showing how temperature variations affect the visibility of the HOM interference. The correlation functions are normalized by the maximum coincidences of the side peaks. b) In the upper row, the probability of measuring the coincidences between the distinct outputs of the beamsplitter, $P_{0}$, is shown as a function of temperature. In the lower row, the visibility of HOM interference is plotted against temperature. For each row, data for the biexciton and exciton are displayed on the left and right, respectively. }
    \label{fig:HOM_vs_temp}
\end{figure}


\section{Results}

We investigated the experimentally measured correlation functions of HOM interference at each sampled temperature, as shown in Fig.~\ref{fig:HOM_vs_temp}a). To quantitatively analyze these correlation functions, we calculated the probability \( P_{0} \) of measuring coincidences between the beamsplitter outputs, along with the corresponding visibilities.  

The probability \( P_{0} \) is given by $P_{0} = A_{c}/(A_{l} + A_{r})$, where \( A_i \) represents the area of a peak, and the subscripts \( c \), \( l \), and \( r \) denote the central, left, and right peaks, respectively. The two side peaks arise when the photons do not impinge on the beamsplitter simultaneously, resulting in coincidences recorded in the absence of HOM interference. In contrast, the central peak, where coincidences are recorded in the presence of HOM interference, is suppressed compared to the side peaks.  

The visibility of HOM interference is defined as $(P_{\infty} - P_{0})/(P_{\infty} + P_{0})$, where \( P_{\infty} \) represents the probability of coincidences when fully distinguishable photons impinge on the beamsplitter~\cite{paul_interference_1986,kaltenbaek_experimental_2006}. Since the photons are fully distinguishable in this case, \( P_{\infty} \) is equal to 0.5. The correlation functions shown in Fig.~\ref{fig:HOM_vs_temp}a) are normalized to the average of the maximum coincidences of the two side peaks.  

To calculate $P_0$ and visibility, the measured coincidences were summed over 141 bins for each peak, with a bin size of 16 ps. $P_{0}$ and the visibility as a function of temperature are shown in Fig. \ref{fig:HOM_vs_temp} b). As the temperature increases from 4.8 K to 26.9 K, we observe that $P_{0}$ for biexciton photons increases from 0.288(3) to 0.373(3) and for exciton photons from 0.335(3) to 0.423(4). Correspondingly, the visibility decreases from 0.270(8) to 0.146(9) for the biexciton photons and from 0.198(8) to 0.083(8) for the exciton photons. The decrease in HOM visibility is attributed to increased phonon-induced dephasing at higher temperatures. This phonon-induced dephasing reduces the coherence time of the photons, thereby decreasing the visibility of HOM interference\cite{thoma_exploring_2016}.

\begin{figure}[ht!]
\centering
\includegraphics[width=0.7\linewidth]{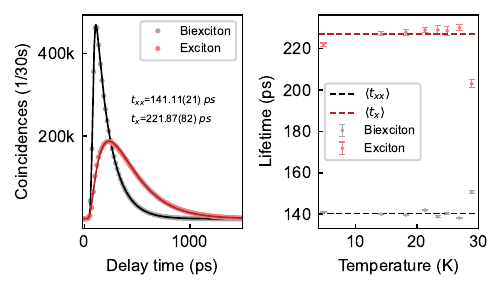}
\caption{Time-resolved photoluminescence measurements for the biexciton and exciton under two-photon resonant excitation at 4.8 K are shown on the left, alongside the extracted lifetimes as a function of temperature on the right. In the left plot, the solid lines represent fitted curves, while dotted lines indicate the measured data. At 4.8 K, the extracted lifetime of the biexciton is 141.11(21) ps, and that of the exciton is 221.9(8) ps. In the right-hand plot, the horizontal dashed lines mark the lifetimes used for the simulation of $G^{(2)}_{HOM}(\tau)$ via the sensor method for both the biexciton and exciton. These lifetimes are 140.40(16) ps for the biexciton and 227.2(6) ps for the exciton.}
\label{fig:lifetime}
\end{figure}


At each sampled temperature, we observed that the lifetimes remained stable, indicating that the temporal correlation between the biexciton and exciton photons varied only marginally. Since the temporal correlation between the pair of emitted photons remains unchanged with stable lifetimes at different temperatures, as shown in Fig. \ref{fig:lifetime}, we conclude that the reduction in HOM interference visibility is solely due to phonon-induced dephasing. We stopped changing the temperature above 26.8 K, where the photon count rates began to decrease significantly and the lifetimes no longer remained constant. The lifetimes of the biexciton and exciton change by 11\% and 7\%, respectively, from 26.8 K to 28.8 K, causing the temporal correlation to differ from what it was before 26.8 K, as shown in Fig. \ref{fig:lifetime}. This drastic change in both the spectrum and the lifetimes is likely due to the thermally induced carrier escape from the quantum dot and an enhancement of non-radiative recombination of electron-hole pairs\cite{heitz_temperature_1999,lubyshev_exciton_1996,zhang_dynamic_2000}. The overall trend of the spectra and lifetimes as a function of temperature is consistent with Ref. \cite{heitz_temperature_1999}.
\begin{figure}[h!t]
    \centering 
    \includegraphics[width=0.7\linewidth]{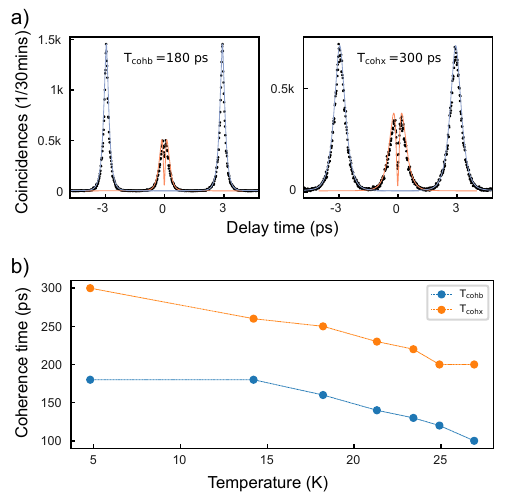}
    \caption{a) Measured correlation functions of HOM interference for the biexciton (left) and exciton (right) photons at 4.8 K are shown along with closely matched theoretical simulation of $G^{(2)}_{HOM}(\tau)$ using the sensor method. This simulation is based on the measured lifetimes and coherence times of 180 ps and 300 ps for the biexciton and exciton photons, respectively, which produce closely matched correlation functions. Dotted lines represent the measured correlation functions, while solid lines show the theoretically predicted $G^{(2)}_{HOM}(\tau)$ in the presence (orange) and absence (blue) of HOM interference. The simulated correlation functions are convolved with the system response functions. b) The estimated coherence times for the biexciton and exciton photons, extracted as described in a), are presented over the sampled temperatures. }
    \label{fig:coh}
\end{figure}


To estimate the coherence times, we simulated
$G^{(2)}_{HOM}(\tau)$ for a range of coherence times with experimentally determined lifetimes. The coherence times were varied from 50 to 200 ps for the biexciton and 200 to 300 ps for the exciton, with a step size of 10 ps. The coherence times of the biexciton and exciton photons were extracted by identifying which of the simulated $G^{(2)}_{HOM}(\tau)$, convolved with the system response, best represents the correlation function measured experimentally at each temperature. For example, Fig. \ref{fig:coh}a shows the experimentally measured correlation functions at 4.8 K, along with their theoretical predictions. At this temperature, the extracted coherence times ($T_{2}$) are 180(10) ps for the biexciton and 300(10) ps for the exciton photons. Based on these values, we deduced the dephasing times to be 510(65) ps for the biexciton and 883(76) ps for the exciton. Here, the dephasing time is calculated using the relation $\frac{1}{T_{2}}=\frac{1}{2T_{1}}+\frac{1}{T^{*}_{2}}$, where $T_{1}$ and $T^{*}_{2}$ refer to the lifetime of a given state and the dephasing time, respectively.
To investigate how phonon-induced dephasing affects coherence times and consequently HOM interference visibility, we extracted coherence times as a function of sampled temperature, as shown in Fig. \ref{fig:coh}b). We conclude that phonon-induced dephasing primarily reduces the coherence times of both the biexciton and exciton, thereby decreasing HOM interference visibility. This effect is independent of the temporal correlation caused by cascaded emission, as the lifetimes remain largely unchanged, as shown in Fig. \ref{fig:lifetime}.

\section{Conclusion}
In this experiment, we examined how the phonon-induced dephasing affects the coherence times of emitted photons. To achieve this, we systematically varied phonon-induced dephasing while maintaining the temporal correlation in the cascaded emission.  We determined coherence times of single photon pairs in the cascaded emission by investigating the correlation functions of HOM interferenc. For this, we used SNSPDs with high temporal resolution and a sensor method to model correlation functions based on measured lifetimes and preselected coherence times. 

To our knowledge, this is the first study to investigate the coherence times of both biexciton and exciton photons as a function of temperature under two-photon resonant excitation. We mapped the visibility of HOM interference across sampled temperatures and extracted the corresponding coherence times. The HOM visibilities of the biexciton and exciton photons decreased by 46.0\% and 58.2\%, respectively, between 4.8 K and 26.9 K. Correspondingly, the coherence times decreased by 44.4\% and 33.3\% for the biexciton and exciton photons, respectively. From this, we find that the HOM visibility of biexciton photons is more susceptible to temperature variation than that of the exciton photons. In contrast, the coherence time of the exciton photons is less significantly influenced by the temperature change, indicating that the exciton emission is less sensitive to phonon-induced dephasing.

Our findings emphasize the importance of the engineering of phonon-induced dephasing to improve HOM visibility. This highlights the need to engineer material properties to minimize dephasing and enhance the performance of quantum dot-based photon pair sources.

\section{Funding}
CAPES/STINT, grant No. 88887.646229/2021-01; Swedish Research Council grant 2021-04494; Carl Tryggers Stiftelse CTS 24:3526.

\section{Acknowledgment}
J. L. and C. S were supported by the Knut \& Alice Wallenberg Foundation (through the Wallenberg Centre for Quantum Technology (WACQT)). A.P. would like to acknowledge the Swedish Research Council (grant 2021-04494) and Carl Tryggers Stiftelse (grant CTS 24:3526).

\bibliography{references}

\end{document}